# Influence of Inter-Layer Exchanges on Vorticity-Aligned Colloidal String Assembly in a Simple Shear Flow

Xinliang Xu[1,2], Stuart A. Rice[3,*], Aaron R. Dinner[3,*]

[1]Department of Chemistry, MIT, Cambridge, MA 02139, USA

[2]Pillar of Engineering Product Development, Singapore University of Technology and Design, 138682, Singapore

[3]James Franck Institute and Department of Chemistry, University of Chicago, Chicago, Illinois 60637, United States

**Corresponding Authors**

[*]Email: sarice@uchicago.edu; Phone: +1 773 702 7199; Fax: +1 773 702 5863 (S.A.R.).

[*]Email: dinner@uchicago.edu; Phone: +1 773 702 2330; Fax: +1 773 702 4180 (A.R.D.).

**Abstract**

Hard spheres in Newtonian fluids serve as paradigms for Non-Newtonian materials phenomena exhibited by colloidal suspensions. A recent experimental study (Cheng et al. **2011** *Science*, *333*, 1276) showed that upon application of shear to such a system, the particles form string-like structures aligned in the vorticity direction. We explore the mechanism underlying this out-of-equilibrium organization with Steered Transition Path Sampling, which allows us to bias the Brownian contribution to rotations of close pairs of particles and alter the dynamics of the suspension in a controlled fashion. Our results show a strong correlation between the string structures and the rotation dynamics. Specifically, the simulations show that accelerating the rotations of close pairs of particles, not increasing their frequency, favors formation of the strings. This insight delineates the roles of hydrodynamics, Brownian motion, and particle packing, and, in turn, informs design strategies for controlling the assembly of large-scale particle structures.

**Table of Contents Graphic**

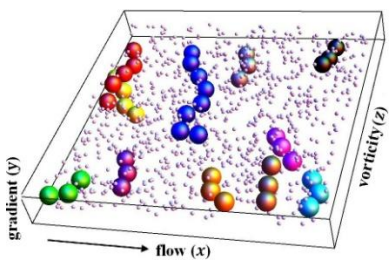

Colloidal dispersions exhibit diverse mechanical[1,2] and transport[3,4] properties that find a wide range of applications in emerging technologies[5] and in consumer products[6]. These useful properties of colloidal dispersions are related to the microstructure of the dispersion—the spatial organization of the colloidal particles. For many applications, it is of great value to be able to switch between different microstructures by means of an external field, e.g., a fluid flow.

When a colloidal suspension is subjected to flow, the microstructure rearranges as the suspension is driven out of equilibrium[7,8]. Depending on the complex interplay between hydrodynamics, Brownian motion, and particle packing, various structures can be accessed[9-13]. We expect the out-of-equilibrium structure generated to depend on the direct colloid-colloid interaction (e.g., hard sphere repulsion versus Janus sphere interaction[14]), the particle shape[15], and the character of the particle surface. Understanding how these particle characteristics, and their interaction with the external field, generate particular microstructures will permit design of desired structures by manipulation of particle properties. While many experimental and theoretical efforts have attempted to investigate the intimate coupling between the flow and the flow-induced microstructure[16], a comprehensive understanding is still missing.

A hard sphere colloidal suspension that is confined between parallel plates and subjected to a simple shear flow captures many non-Newtonian behaviors and thus serves as a simple system for the study of these complex phenomena (see ref. 17 and references therein). In a recent experimental analysis of such a system[17], Cheng et al. observed shear-dependent assembly of log-rolling colloidal strings within each layer of colloidal particles parallel to the confining plates. We studied this string structure using Stokesian Dynamics simulations, and found evidence that particle exchanges between layers are important[18]. We further proposed that one mode of particle exchanges between layers—the rotation of close pairs of spheres induced by the shear flow—is responsible for the observed formation of strings. However, the shear rates and the particle exchange rates in the simulations could not be varied independently to test the proposed mechanism directly. Advances in simulation algorithms now make this more detailed exploration possible. Using a recently developed method for sampling rare events, Steered Transition Path Sampling[19], in combination with Stokesian Dynamics simulations[20, 21], we now identify the precise mechanism that leads to the string structure. We find that the time scale, rather than the number, of close pair rotations is the key factor in string formation. This distinction is important as different manipulations of the particle shape and packing, as well as the fluid properties, will favor one or the other and, in turn, different structures.

In this study we use Stokesian Dynamics to simulate the dynamics of a monodisperse suspension of $N = 100$ hard spheres of unit radius. Our simulations use the methodology developed by Brady and coworkers[21, 22], where the influence of hydrodynamic interactions on the particle motions is incorporated through the resistance tensor evaluated for the configuration at the beginning of each time step of length $\Delta t$. The centers of the spheres are restricted to a rectangular box spanning $0 < x < 25$, $0 < z < 25$ and $-1.3 < y < 1.3$ (Figure 1A). We apply periodic boundary conditions in the *x* and *z* directions and an external confining potential along the *y* direction that is 0 for $|y| \leq y_0 = 1.1$ and $U_y = 200 k_B T (|y| - y_0)^2$ otherwise. The suspension is subject to a simple shear flow in the *x* direction with shear gradient in the *y* direction and the vorticity in the *z* direction. The sheared hard sphere fluid is characterized by two dimensionless parameters, the packing fraction $\phi = N(4\pi a^3)/3V$ and the Péclet number $Pe \equiv 6\pi \eta_0 a^3 \dot{\gamma}/k_B T$, where $a$ is the radius of a sphere, $V$ is the volume of the system, $\eta_0$ is the viscosity of the supporting liquid, and $\dot{\gamma}$ is the shear rate. The Peclet number measures the magnitude of particle motion induced by shear relative to that associated with thermal motion.

The unit time of the simulation $t_0$ is defined in terms of the diffusive time scale $6\pi\eta_0 a^3/k_B T$ when $Pe < 1$ and the inverse shear rate $1/\dot{\gamma}$ when $Pe > 1$.

Simulations were carried out for simple shear flows in a system with packing fraction 0.22 for $Pe = 10, 30, 50$ and $70$. To evaluate the flow-induced rearrangements for each $Pe$, 100 independent simulation runs were initialized using a Monte Carlo method. Then in each Stokesian Dynamics run the system was equilibrated for 25,000 time steps of $\Delta t = 10^{-3}$ before applying shear, i.e., at $Pe = 0$. Another 50,000 time steps were taken after the introduction of the shear flow to allow the system to come to a non-equilibrium steady state, after which 100,000 time steps were taken for analysis of averaged properties.

Our results reproduce earlier observations of shear induced layering in the non-equilibrium steady state[18], characterized qualitatively by the peaks in the density profile $\rho(y)$ along the shear gradient direction *y* (Figure 2A). Within each layer (*xz* plane), the probability of finding neighboring spheres (separation $r < 2.1a$) with relative alignment angle $\theta$ with respect to the flow direction *x* (Figure 2B, Inset) was investigated using a normalized angular probability function $P(\theta)$ satisfying $\int_0^\pi P(\theta)d\theta = \pi$. Our results show vorticity-aligned strings[18], characterized qualitatively by the strong anisotropic distribution displayed in $P(\theta)$, favoring alignment in the *z* direction with $\theta = \pi/2$ (Figure 2B). To quantitatively investigate this anisotropy, we introduce a scalar parameter $Q_A \equiv -\int_0^\pi P(\theta)\cos 2\theta\, d\theta$, where the amplitude quantifies the level of anisotropy and the sign shows whether alignment along *z* is preferred (positive) or alignment along *x* is preferred (negative). The dependence of the average of $Q_A$ in the non-equilibrium steady state on $Pe$ is displayed in Figure 2C. Our results show an increase of anisotropy as $Pe$ increases, in agreement with earlier observations[17, 18].

We previously showed that introduction of an external repulsive potential between layers, which reduces particle exchanges between neighboring layers, can lead to partial or total removal of the vorticity-aligned strings[18]. The strong correlation between the string structure and the repulsive potential between layers suggested that particle exchanges between neighboring layers are important in the formation of the strings. Particle migration driven by flow has been an active field of research[23]. Here we focus attention on one mode of particle exchanges between neighboring layers, namely, rotations of close pairs of spheres in the *xy* plane, where two spheres *i* and *j* are labeled as a close pair in the *xy* plane when the 3D separation $r_{ij} < 2.1a$ and $\frac{|z_j - z_i|}{r_{ij}} < \varepsilon$, where $\varepsilon$ is a cutoff for relative out-of-plane displacement. We take $\varepsilon = 0.8$, which is a rather generous criterion, to improve the statistics by including occasional large *z* separations during the rotational dynamics. However, we find that the results obtained are insensitive to variations between $\varepsilon = 0.5$ and $\varepsilon = 0.8$. The rotation of a close pair in the *xy* plane is characterized by $\psi$, the angle between the *x*-axis and the relative alignment projected in the *xy* plane (Figure 1B).

When subject to a simple shear, the flow-induced motion for each sphere in a close pair in the *xy* plane with respect to the center of mass, $\vec{v}_i$ and $\vec{v}_j$, can be decoupled into a radial part and an angular part (Figure 1C). The changes in microstructure caused by the interplay between the radial stretching (or compressing) and the nearest neighbor lubrication force has been discussed in many earlier studies[24, 25]. Here we demonstrate that the vorticity-aligned string structure is strongly correlated to the angular motion, which is manifest as rotation in the *xy* plane. For each close pair (denoted by index *n*), the angular displacement over a time duration $l_t$, i.e., $\Delta\psi_n(l_t) = \psi_n(t + l_t) - \psi_n(t)$, contains contributions from both the flow-induced rotation with angular velocity $\omega(\psi_n) = \dot{\gamma}\sin^2\psi_n \sim Pe$ and the Brownian motion, which is independent of $\psi_n$. The contribution from the Brownian motion can be obtained by removal of the flow-induced rotational contribution: $\delta\psi_n(l_t) = \Delta\psi_n - \dot{\gamma}\sin^2\psi_n \times l_t$. Using 200,000 non-equilibrium steady state configurations for the suspension with $Pe = 30$, we found that $\delta\psi_n(l_t)$ can be well described by a diffusion process with $\delta\psi_n(l_t) = \mathcal{N}(0, 2D_{\psi_n}l_t)$, where $\mathcal{N}(\mu, \sigma^2)$ denotes a normal distribution with an average of $\mu$ and a standard deviation of $\sigma$, and $D_{\psi_n} = 0.009\ radian^2/t_0$ characterizes the angular diffusion coefficient. Defining $\delta\psi(l_t) = \sum_n \delta\psi_n(l_t)$, where the sum goes over all the close pairs within a specific range of $\psi$, we construct a histogram of values of $\delta\psi$ (Figure 3A) at time interval $l_t = 0.05 \times t_0$ and $\psi < \pi/6$, where the Brownian motion plays a more important role since the rotational contribution $\omega(\psi) \sim \psi^2$ at small $\psi$ is small. This distribution delineates the amplitude of thermal fluctuations, and is well described by a Gaussian function with an average value of $10^{-3}$ and a standard deviation of 0.09, both in units of *radians*.

If we are able to harvest a sufficient number of independent trajectories, each consisting of many successive segments of length $l_t = 0.05 \times t_0$ in the large $\delta\psi$ tail of this distribution, we can evaluate the microstructure associated with the configurations along those trajectories and compare the string structure found with the string structure observed in the non-equilibrium steady state. Doing so is challenging because these trajectories are rare. We overcome this difficulty using Steered Transition Path Sampling[19]. This numerical technique allows control over the sampling of currents of dynamic events without otherwise distorting the dynamics. The essential idea is that a small probability can be decomposed into a product of probabilities near unity by breaking a process of interest into successive shorter trajectory segments of length $l_t$ that make incremental progress. Starting from the same initial configuration we generate a large number, $N_T$, of short trajectory segments according to the original dynamics. From these segments, we obtain an estimate of the probability that segments satisfy a progress constraint: $P \sim N^+/N_T$, where $N^+$ is the number of segments satisfying the constraint. By introducing a bias threshold $P_{Bias}$, we select among the $N_T$ segments with uniform probability when $P > P_{Bias}$; otherwise we select one among the $N^+$ segments with probability $P_{Bias}$. We then correct for this bias in calculating averages with a reweighting factor (see ref. 19 for details of the procedure).

For each of the 100 independent Stokesian Dynamics simulation runs with $Pe = 30$, we choose the configuration 70,000 steps after the onset of the shear motion to initialize a Steered Transition Path Sampling simulation. In each Steered Transition Path Sampling simulation, a trajectory of 20,000 time steps of $\Delta t = 10^{-3}$ is built up by linking 400 successive short trajectory segments, each with length $l_t = 50$ time steps. To obtain any one of such short trajectory segments, we generate many stochastic realizations until we obtain one satisfying $\delta \psi > 0.03$. Then we select among these trajectory segments as described above with $P_{Bias} = 0.95$. Using this method, we can readily obtain trajectories in which close pair rotations are assisted by biased Brownian motions and are thus much faster while maintaining $Pe = 30$. In this way, we can independently explore the applied shear rate and the rate of particle exchanges between neighboring layers caused by close pair rotations. Along the trajectories obtained from the Steered Transition Path Sampling simulations the configurations and the microstructure obtained show a strong enhancement of the vorticity-aligned string structure (Figure 3B). This result explicitly demonstrates that particle exchanges due to close pair rotations in the suspension underlie the string structure.

While determining the time scale of each inter-layer particle exchange due to close pair rotation $\tau_{rot}$ is a complex first-time passage problem, Steered Transition Path Sampling allows us to decouple $\tau_{rot}$ from the applied shear rate simply by selecting trajectories in which the Brownian motion favors rapid rotation. The fact that we can control the assembly of the vorticity-aligned string structure by varying $\tau_{rot}$ strongly supports the idea that the observed anisotropic microstructure results from the competition between the intra-layer relaxation back to the isotropic structure and the inter-layer particle exchanges that give birth to the strings. This competition can be described by one dimensionless unit, the ratio between $\tau_{rel}$ the time scale of the intra-layer relaxation and $\tau_{rot}$ the time scale of the inter-layer particle exchanges due to close pair rotations.

It is important to note that it is the time scale, not the number of the inter-layer particle exchanges due to close pair rotations observed in a given interval of time, that is important in the formation of the strings. By using Steered Transition Path Sampling, we have sampled rare trajectories where the number of close pair rotations in each time interval of $2t_0$ is one standard deviation higher than the average. The resulting configurations along these rare trajectories show no notable changes in the string structure (Figure 3B). This result can be qualitatively explained by analogy with a two-state chemical reaction. Within each layer, pairs aligned in the *x* direction and in the *z* direction can be viewed as two metastable states, denoted as *X* and *Z*, with relative populations $n_x$ and $n_z$, respectively. Escape from these metastable states occurs with forward rate $k_{X \to Z} \sim \tau_{rot}^{-1}$ and backward rate $k_{Z \to X} \sim \tau_{rel}^{-1}$. The string structure, characterized by the anisotropy favoring the *z* direction or $\langle n_z \rangle / \langle n_x \rangle > 1$, is controlled by $\langle n_z \rangle / \langle n_x \rangle = k_{X \to Z}/k_{Z \to X} \sim \tau_{rel}/\tau_{rot}$, the ratio of timescales according to reaction rate theory. On the other hand, our sampled trajectories associated with a larger number of rotation events within the same period of time are related to a larger forward flux characterized by $k_{X \to Z} n_x > \langle k_{X \to Z} n_x \rangle$. Such a flux can be achieved with either $k_{X \to Z} > \langle k_{X \to Z} \rangle$ or $n_x > \langle n_x \rangle$. Since $k_{X \to Z} > \langle k_{X \to Z} \rangle$ enhances the resulting anisotropy $n_z/n_x$ while $n_x > \langle n_x \rangle$ weakens it, by mixing these two types of events without differentiation our sampled trajectories are unable to change the resulting string structures. Making the distinction between the time scale and number of exchanges required the simulations reported in the present study.

Given the apparent trend observed in the recent experiment, namely, that stronger string structures are accompanied by simultaneous formation of stronger layering[17, 18], it is natural to ask if particle layering plays an important role in shaping the vorticity-aligned strings. Our determination of the close pair rotations as the underlying mechanism implies that layer ordering of the suspension is not essential to the formation of the string structure, as there exists no explicit relation between the close pair rotation dynamics and the layer ordering of the suspension. To strengthen this conclusion, we define a scalar parameter $Q_L = \sum_{i=1}^{N} |y_i|/N$ that characterizes the layer ordering for each configuration. Using 200,000 steady-state configurations for the suspension with $Pe = 50$, we construct a histogram of values of $Q_L$ (figure 4A), which is well represented by a Gaussian function with mean of 0.759 and standard deviation of 0.028. Our results show that the string structure, characterized by $Q_A$ associated with configurations at each value of $Q_L$, has a very weak dependence on $Q_L$ (figure 4B). This weak relation between the string structure and particle layering can be better visualized by dividing those 200,000 steady-state configurations into three groups: $Q_L > 0.787$, $0.787 > Q_L > 0.731$ and $Q_L < 0.731$, for which the corresponding density profiles show notable differences in peak heights (Figure 4C). The angular probability functions $P(\theta)$ associated with those configurations within each group show very small variances (figure 4D), supporting our conclusion that layer ordering is not important to the formation of the vorticity-aligned string structure.

In summary, we have used numerical simulation methods to investigate the recently experimentally observed vorticity-aligned string structure in a confined colloidal suspension under simple shear flow. By sampling the rare trajectories with different close pair rotation speeds while keeping the shear rate fixed we are able to control the level of the anisotropy in the system, leading us to the conclusion that the time scale for each inter-layer particle exchange caused by close pair rotation underlies the observed string structure. Drawing on this conclusion, we anticipate that particle pairs with different (hydrodynamic) interactions will rearrange differently upon shear as compared to the hard spherical particles studied here. For example, consider two separated particles, each of which rotates around its own center of mass in the $xy$ plane when subjected to a simple shear flow. If the rotational degree of freedom is repressed (or enhanced) by the pair interactions when the two particles form a close pair (this can be realized by rough surfaces or different shapes other than spheres), the rotational angular momentum of each particle will be transformed into the rotation of the pair around the center of mass of the pair in the $xy$ plane, and thus change the rotation dynamics discussed in this work. As a result, one can have particles with prescribed interactions that, when subject to shear, will rearrange and form a structure by design.

**Acknowledgements**

The research work by X.L.X. is supported by a research fellowship by Singapore University of Technology and Design. S.A.R. and A.R.D. acknowledge the financial and central facilities assistance of the University of Chicago Materials Research Science and Engineering Center, supported by National Science Foundation Grant DMR-MRSEC 0820054.

**Additional information**

The authors declare no competing financial interest.

**Figure Captions**

**Figure 1. Description of the suspension.** (**A**) Illustration of the coordinate system. (**B**) Illustration of a counter-clockwise rotation of a close pair of particles as projected in the *xy* plane at an earlier stage (circles with solid lines) and at a later stage (circles with dashed lines). (**C**) The flow-induced motion for each sphere in such a pair with respect to the center of mass, $v_i$ for particle $i$, can be decomposed into a radial part and an angular part $v_{i\psi}$.

**Figure 2. Flow-induced rearrangements.** Illustration of (**A**) the density profile $\rho(y)$ and (**B**) the angular distribution $P(\theta)$, for a suspension at non-equilibrium steady state with $Pe = 50$. Quantitative description of the anisotropy shown in $P(\theta)$ can be measured by $Q_A$ as a function of the flow strength $Pe$, as illustrated in (**C**). Each data point in (**C**) is the average of 100 independent simulations, and the bars show standard deviations.

**Figure 3. Enhanced vorticity-aligned string structure by Steered Transition Path Sampling.** (**A**) Histogram of $\delta\psi$ for 200,000 steady-state configurations of the suspension with $Pe = 30$, fitted to a Gaussian function (line). (**B**) Quantitative description of the vorticity-aligned strings $Q_A(t)$ at time $t$ for suspensions with $Pe = 30$, analyzed from configurations obtained from normal Stokesian Dynamics simulations (black squares), Steered Transition Path Sampling simulations with $\delta\psi > 0.03$ (red circles), and Steered Transition Path Sampling simulations with increased number of rotations per time duration of $2t_0$ (blue diamonds). The vertical dashed lines at $t = t_1$ and $t = t_2$ correspond to the onset of the shear motion and the onset of the Steered Transition Path Sampling, respectively. The horizontal dashed line at $Q_A = Q^*$ illustrates the average of $Q_A$ in the non-equilibrium steady state for the suspension with $Pe = 30$.

**Figure 4. String structure and layer ordering.** (**A**) Histogram of $Q_L$ for 200,000 steady-state configurations of the suspension with $Pe = 50$, fitted to a Gaussian function (line). These configurations are divided into three groups: large $Q_L$ group (blue), medium $Q_L$ group (green) and small $Q_L$ group (red). (**B**) The string structure, characterized by $Q_A$, associated with configurations at each value of $Q_L$. Illustration of (**C**) the Density profiles $\rho(y)$; and (**D**) the angular distributions $P(\theta)$, obtained from configurations in large (blue), medium (green) and small (red) $Q_L$ group. Each data point in (**B**) is the average of 100 independent simulations, and the bars show standard deviations.

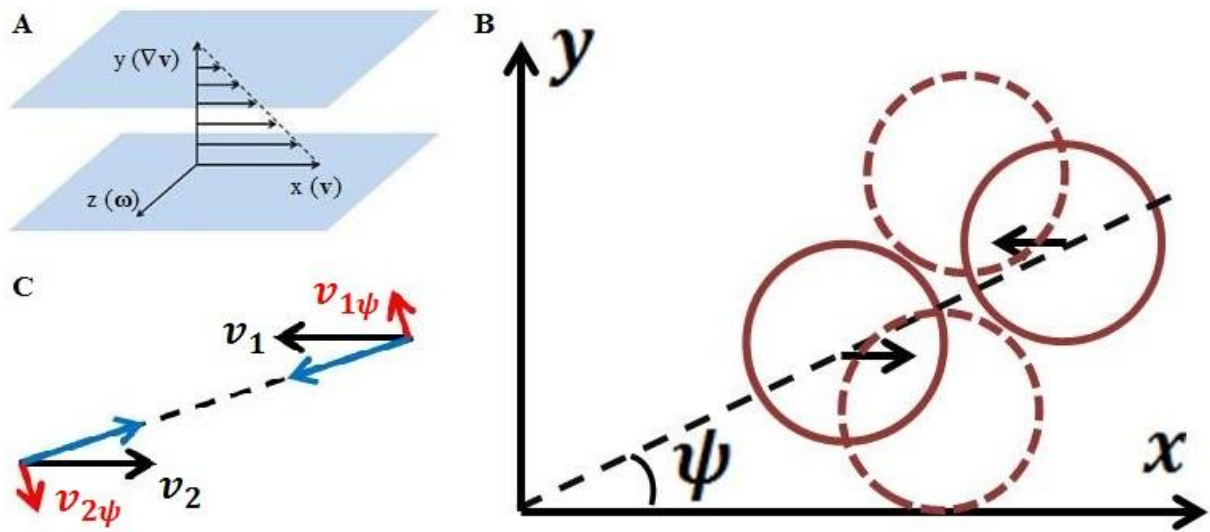

Figure 1

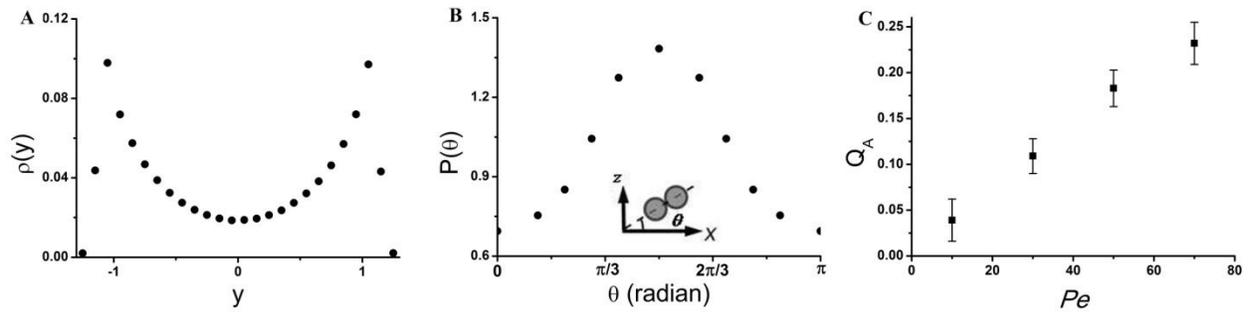

Figure 2.

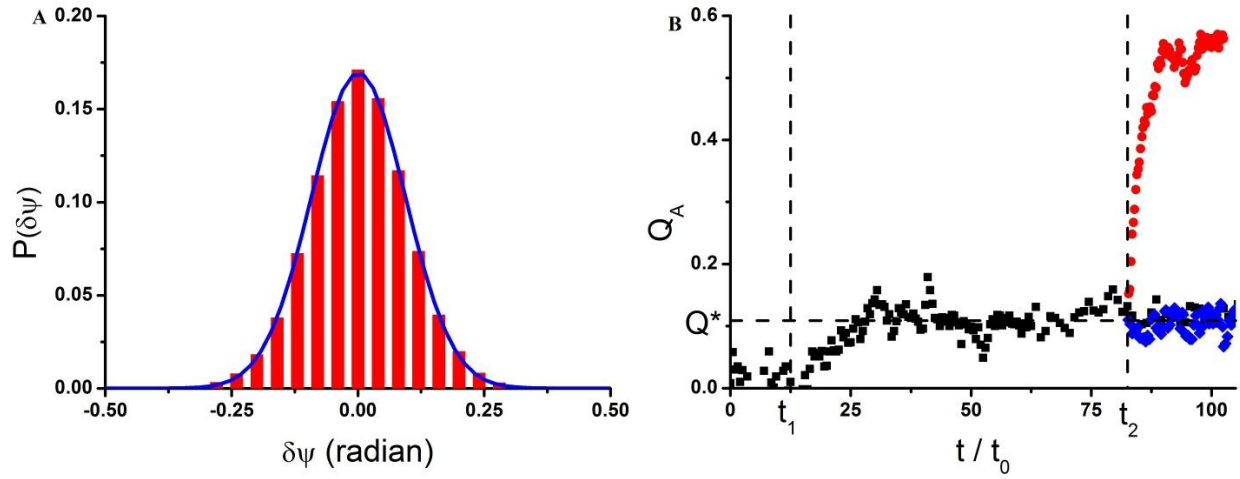

Figure 3.

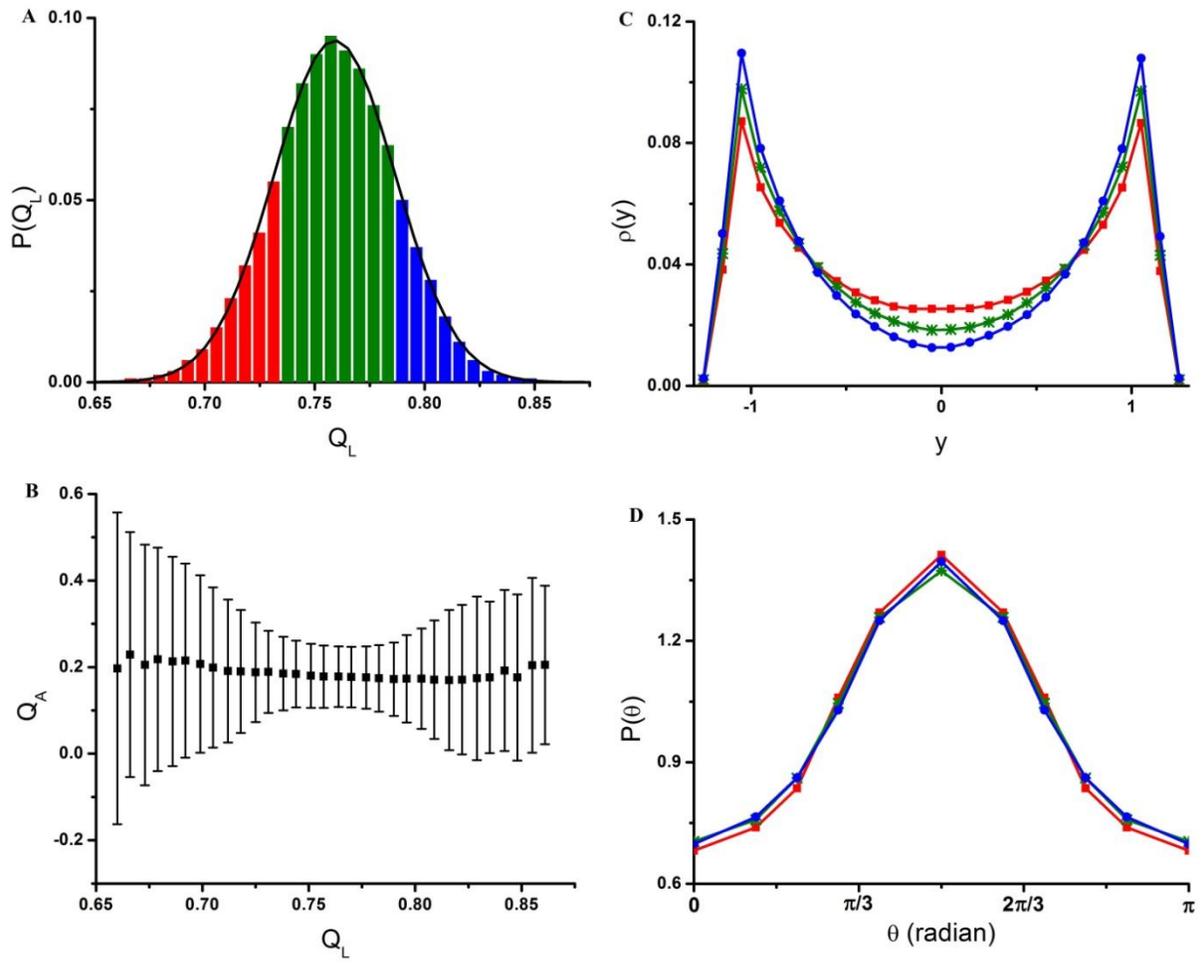

Figure 4.